\documentclass[%
 reprint,
superscriptaddress,
groupedaddress,
showpacs,preprintnumbers,
nofootinbib,
nobibnotes,
 amsmath,amssymb,onecolumn,
prb,
]{revtex4-1}

\usepackage{graphicx}
\usepackage{dcolumn}
\usepackage{bm}
\usepackage{amssymb}   
\usepackage{amsmath}
\usepackage{color}
\usepackage[normalem]{ulem}
\usepackage{siunitx}
\usepackage{mathrsfs}
\usepackage{tikz}
\usepackage{amsfonts}
\usepackage{kantlipsum}

\begin{document}


\title{Multiplexing-oriented plasmon-MoS2 hybrid metasurfaces driven by selectable suppressed and enhanced nonlinear quasi bound states in the continuum }%

\author{Qun Ren$^{1}$, Feng Feng$^{1\ast}$, Xiang Yao$^{1}$, Quan Xu$^{1}$, Zhihao Lan$^{2}$, Jianwei You$^{2}$, Xiaofei Xiao$^{3}$, Wei E. I. Sha$^{4\ast}$, Ming Xin$^{1}$}

\email{1. School of Electrical and Information Engineering, Tianjin University, Tianjin, 300072, China \\
2. Department of Electronic and Electrical Engineering, University College London, United Kingdom \\
3. Department of Physics, Imperial College London, London, SW72AZ, United Kingdom \\
4. Key Laboratory of Micro-Nano Electronic Devices and Smart Systems of Zhejiang Province, College of Information Science and Electronic Engineering, Zhejiang University, Hangzhou, 310027, China \\
Email: feng_feng@tju.edu.cn, weisha@zju.edu.cn, xinm@tju.edu.cn}

\date{\today}
\begin{abstract}

Rapid progress in nonlinear plasmonic metasurfaces enabled many novel optical characteristics for metasurfaces, with potential applications in frequency metrology\cite{mcr04ol}, timing characterization\cite{xkm17light} and quantum information\cite{mcpl17n}. However, the spectrum of nonlinear optical response was typically determined from the linear optical resonance. In this work, a wavelength-multiplexed nonlinear plasmon-MoS2 hybrid metasurface with suppression phenomenon was proposed, where multiple nonlinear signals could to be simultaneously processed and optionally tuned. A clear physical picture to depict the nonlinear plasmonic bound states in the continuum (BICs) was presented, from the perspective of both classical and quantum approaches. Particularly, beyond the ordinary plasmon-polariton effect, we numerically demonstrated a giant BIC-inspired second-order nonlinear susceptibility $10^{-5}$~$m/V$ of MoS2 in the infrared band. The novelty in our study lies in the presence of a quantum oscillator that can be adopted to both suppress and enhance the nonlinear quasi BICs. This selectable nonlinear BIC-based suppression and enhancement effect can optionally block undesired modes, resulting in narrower linewidth as well as smaller quantum decay rates, which is also promising in slow-light-associated technologies.


\end{abstract}



\maketitle

\section{Introduction}\label{Intro}

Nonlinear metasurface is a promising research orientation for diversifying the optical signal processing functions \cite{mzw19lpr, mnw19prl}. Generally, a nonlinear device exhibits constant nonlinear susceptibility in the frequency range far away from the resonance conditions of the material. Metasurfaces may break this limitation utilizing its design flexibility at the nanoscale level, so as to obtain ultra-high nonlinear susceptibilities in several desired wavelength windows, within a compact physical structure. To achieve this goal, new mechanisms are required to significantly enhance the nonlinear conversion efficiencies, with the help of high nonlinear materials positioned close to a specially designed metasurface.

In recent years, there were several discussions \cite{tx17pr, lr19aom, jrqn18nano, gwx19oe, hh19pr, bqs19lpr, tbc19pra, ysqy20oe, kyk19acs, lei20lpr} on plamson-assisted hybrid metamaterials with and without BICs. On one hand, introducing two-dimensional materials into plasmonics would broaden the sort of materials, which explores more opportunities for plasmonic devices. On the other hand, plasmonic polaritons exhibit huge local field enhancement, which is of great help to the enhancement of optical nonlinearities \cite{jzn20sa,nwd18jo}. However, little attention was paid on the direct modulation of nonlinear optical spectrum. Since the nonlinear optical resonance was typically decided by the linear optical response, new theories are to be explored for more novel functionalities such as individually cancelling particular noise in the nonlinear optical spectrum.

Transition metal dichalcogenides (TMDs), as one of the popular nonlinear optical materials, possess large nonlinear susceptibility due to the strong exciton resonance as well as spatial inversion symmetry breaking, which are the prerequisites for high nonlinear optical conversion efficiency. For the prospects of dramatic light trapping ability and flexible spectrum tunability, combining TMDs with other bulk optical systems such as metal nanostructures are an intelligent choice to customize a specific optical environment for regulating optical performance. Although surface plasmon was used for intensifying light-matter interaction and enhancing optical nonlinearity, the strong inherent radiation loss and dispersive properties of metallic plasmons severely limit the quality (Q) factors of optical resonances \cite{aa18jo}. An elegant solution to suppress the radiation losses and increase Q factors is provided by the notion of BICs. BICs were originally proposed in quantum mechanics as localized eigenstates of single particle whose energy is embedded in the continuous eigenvalue state solutions \cite{fd75pra}. Recently, BICs have attracted decent interest in the field of photonics \cite{xuyi2020pra, ksmak18prl}. An ideal BIC occurs at a given value of the continuous parameter, where one of the radiation channels disappears entirely and the Q factor becomes infinite with vanishing resonance linewidth. In engineering practice, limited by structural size and material inherent loss, BIC modes would collapse to leaky modes with a finite lifetime known as quasi-BICs (supercavity modes). Most importantly, when TMD is placed at the hot spots of quasi-BICs, the strongly enhanced localized field induced by plasmons could greatly excite the optical nonlinearity of TMDs.

Inspired by the ideas above, here we propose a plasmon-MoS2 hybrid nonlinear metasurface with multiple radiation channels induced by selectable suppressed and enhanced plasmon-polariton resonance and BICs, which enables optional high nonlinear conversion efficiency at several wavelength windows. Starting from a principle for lowering radiation loss and enhancing optical resonance, we demonstrate an approach for electively controlling light-trapping resonances governed by BICs. Exploiting either in-plane or out-of-plane symmetry breaking in a complex unit cell, the BIC-associated leaky modes with high Q factors can be excited, which introduce large energy concentration and thus substantially boost the optical nonlinearity of the metasurface. We also present the results on smartly engineering of the resonance spectrum, which predicts the tunability of the plasmon-assisted nonlinear metasurfaces.

The remainder of the paper is organized as follows: In the next section, the configurations and material parameters of the proposed metasurfaces are described. Additionally, an intuitive interpretation for the difference between the BIC-based resonances and the ordinary plasmon-polariton resonances is demonstrated, which helps to judge the quasi-BIC modes. In Sec.~III-A, we verify the formation of quasi BIC in our proposed system, via the improved eigenmode analysis for dispersive materials. In Sec.~III-B, we explain the quasi BIC from the perspective of both the classical and quantum mechanisms. In Sec.~III-C, we further compute the enhancement of second harmonic generation (SHG) in our proposed system, utilizing the homogenization approach. In Sec.~III-D, a suggested fabrication process is described, which could help relevant experimental verifications. Finally, the main conclusions are outlined in Sec.~IV.

\section{Principle of Quasi-BIC Resonance Enhancement}\label{prin}

To create large optical resonance with high sensitivity, a designated metasurface array composed of plasmon-polariton resonators is required. Schematic and operating principle of the proposed nonlinear metasurface are illustrated in Fig.~1. The period of unit cell $P_{x}=P_{y} = 500$~nm, the lateral size of gold strips $L_{x} = 400$~nm, with the longitudinal size $L_{y1}=L_{y2} = L = 120$~nm (without symmetry breaking transversely) and the width $w = 50$~nm [Fig.~1(b)]. Here the thickness of monolayer MoS2 is taken as $h_{MoS2}=0.7$~nm [Fig.~1(c)], with the permittivity found in recent works \cite{bfdk5ome, ywj18oe}. The model of gold permittivity is suggested in reference \cite{at08apb}, with the thickness $h_{gold} = 100$~nm [Fig.~1(c)]. A typical planar array of split metal frames is placed on the monolayer MoS2 [Fig.~1(a)] for exciting singularity-associated light-trapping resonances with two approaches of symmetry breaking [transverse and longitudinal cutoff by $\Delta L$ in Fig.~1(d) and $\Delta h$ in Fig.~1(e), respectively). The hybrid structure comprised of metal and MoS2 is excited by a normally incident plane wave polarized transversely to the array symmetry axis. In the first case, as shown in Fig.~1(f), when $\Delta L=60$~nm, a reflection dip appears in the infrared region (red solid curve), compared with the symmetric structures (black dashed curve).

\begin{figure}[!htb]
\centering
\includegraphics[width=0.95\columnwidth]{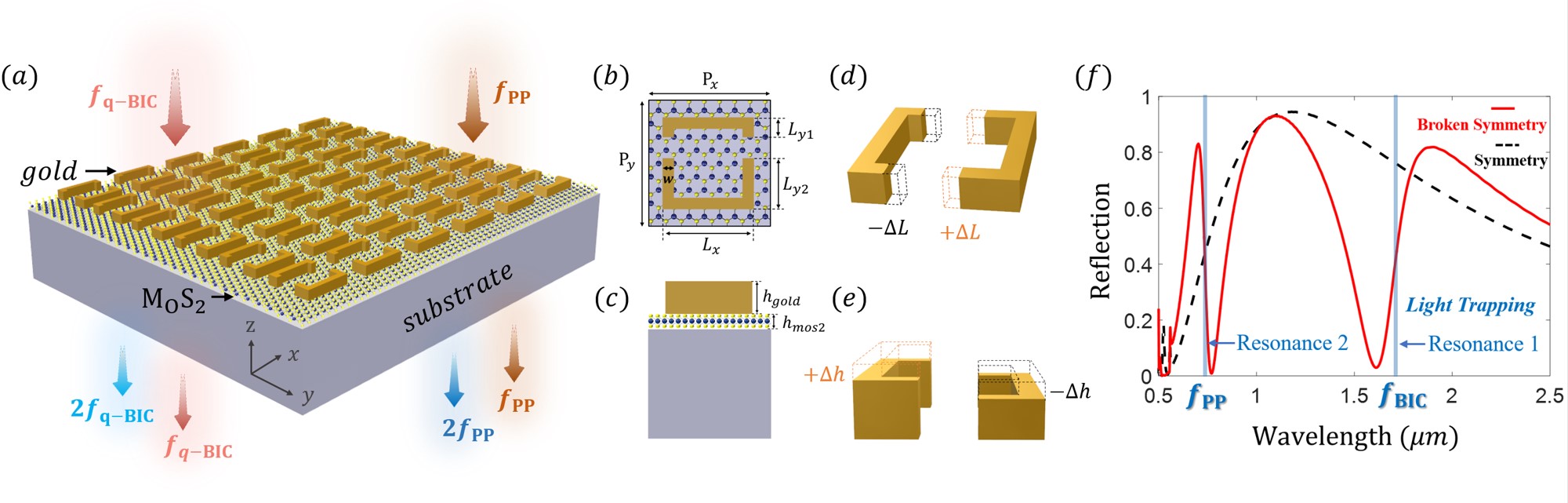}
\caption{(a) Schematics of a nonlinear plasmon-assisted MoS2 metasurfaces with multiple radiation channels orienting towards nonlinear multiplexing. (b) Top view of a unit cell of the targeted metasurface, where the period in $x$ and $y$ direction are $P_{x}$ and $P_{y}$, the length of side are $L_{x}$, $L_{y1}$ and $L_{y2}$, respectively, and $w$ is the width of the metal strips. (c) Front view of a unit cell, where $h_{gold}$ and $h_{MoS2}$ are the thickness of the gold strips and MoS2 layer, respectively. (d,e) Approaches for breaking symmetry in plane and out of plane, respectively. (f) A reflection spectrum of the metasurface. The red line and black dashed line represent the reflection with in-plane symmetry breaking by $\triangle L=60$~nm ($L_{y1}=60$~nm, $L_{y2}=180$~nm) and without the symmetry breaking, respectively. $f_{q-BIC}$ and $f_{PP}$ indicate the light-trapping mode fueled by BIC and a plasmon-polariton resonance mode, respectively.}
\label{schematic}
\end{figure}

The resonance responses of the hybrid meta atoms in this work are studied in the near-infrared wavelength region, which is essential for communication and molecular spectroscopy. The strong near-field enhancement brought by plasmon-polariton resonances can lead to great enhancement of optical nonlinearity, thus realizing multiple high-efficiency nonlinear conversion channels within a compact structure. Due to the inherent limitation of low quality factor imposed by strong metallic dissipation and resistive loss, there is an acute need to gain a deeper insight into the physics and implementation methods of BIC-based linear/nonlinear resonance responses.

First of all, starting from an intuitive interpretation for the difference between the BIC-based resonance mode $f_{q-BIC}$ and the ordinary plasmon-polariton resonance mode $f_{PP}$, the current distributions inside gold strips are plotted for $f_{PP}$ (Fig.~2) and $f_{q-BIC}$ (Fig.~3), respectively. The current density $\textbf{J}$ in gold strips is obtained via the constitute relation $\textbf{J}=\nabla\times \textbf{H} - \partial \textbf{D} / \partial t$ (details for the simulations are elucidated in Appendix A). The normal incident light with $x$-direction polarization excites the (a) in-plane and (b) out-of-plane asymmetric metasurfaces, generating in-phase current oscillations at $f_{PP}$ (Fig.~2) and opposite-phase current oscillations at $f_{q-BIC}$ (Fig.~3), respectively.

\begin{figure}[!htb]
\centering
\includegraphics[width=0.95\columnwidth]{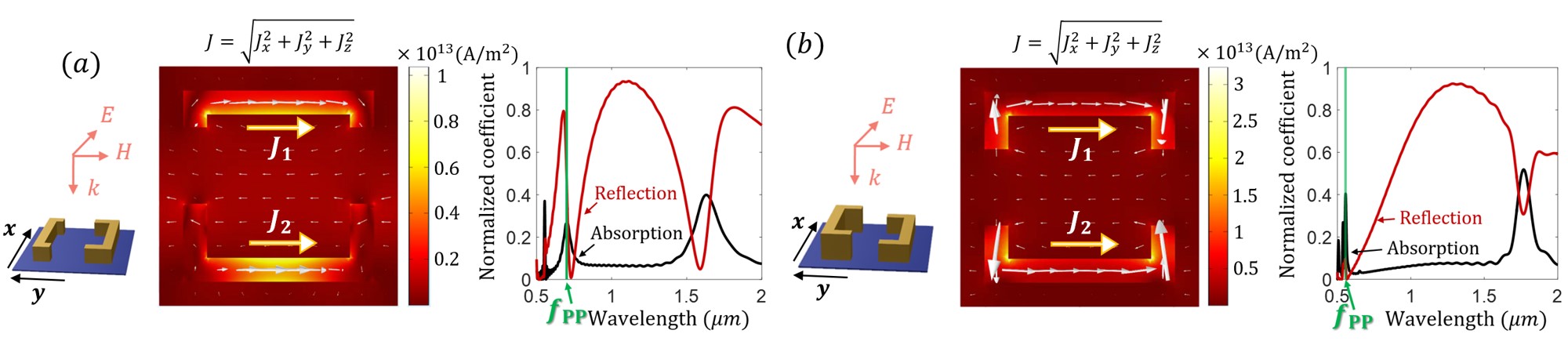}
\caption{Distribution of current density in (a) in-plane and (b) out-of-plane symmetry-broken hybrid unit cells at the plasmon-polariton resonance mode $f_{PP}$. Parameters of the meta atom are (a) $L_{y1}=60$~nm, $L_{y2}=180$~nm, $w = 50$~nm, and $h_{gold} = 100$~nm (b) $L_{y1}=L_{y2}=120$~nm, $w = 50$~nm, $h_{gold1}=120$~nm, and $h_{gold2}= 20$~nm. The incident power $P_{in}=1$~W.}
\label{schematic}
\end{figure}

At the plasmon-polariton resonance mode $f_{PP}$, as shown in Fig.~2, via in-phase collective current oscillations in gold strips, the eigenmodes of the system have strong dipolar momentum and exhibit as bright or radiative modes \cite{eaa18book}, which is the most common way to generate large resonance responses. Due to the strong linear polarization at $f_{PP}$, the corresponding nonlinear conversion efficiency induced by $f_{PP}$ can also be greatly enhanced.

\begin{figure}[!htb]
\centering
\includegraphics[width=0.9\columnwidth]{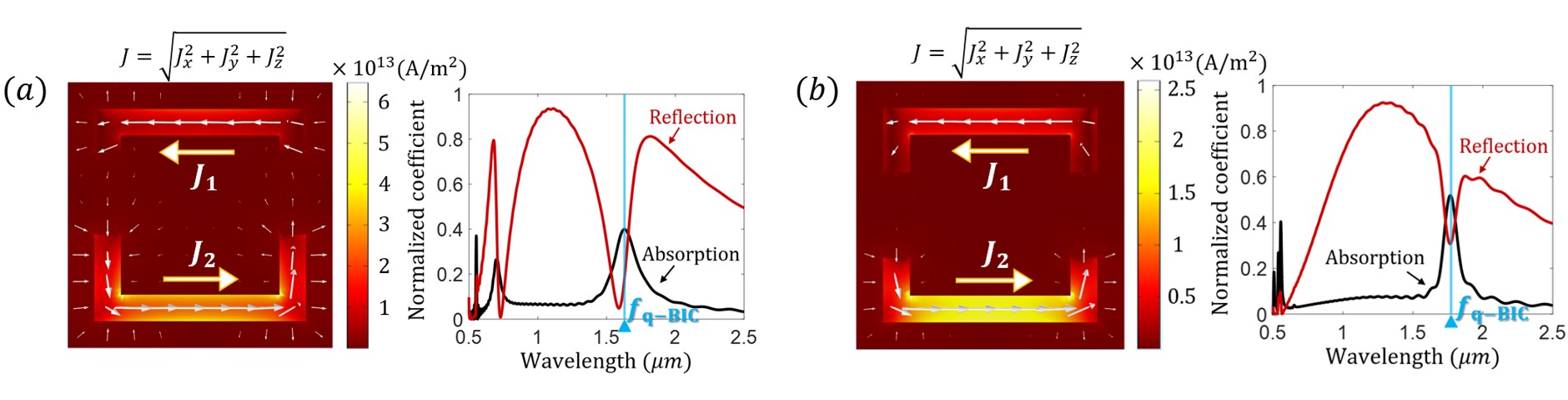}
\caption{Distribution of current density in (a) in-plane and (b) out-of-plane symmetry-broken hybrid unit cells at the light-capturing mode $f_{q-BIC}$ for the same structure described in Fig.~2.}
\label{schematic}
\end{figure}

In contrast, at the BIC-based resonance mode $f_{q-BIC}$, as illustrated in Fig.~3, the current oscillations in gold strips are in opposite directions. For an infinitely periodic arrangement of unit cells, the opposite-phase oscillating currents could be cancelled by pairs, without affecting the radiation in the far-field \cite{nsnv08np}. Therefore the opposite-phase current oscillations produce eigenmodes with low dipolar momentum and little coupling interaction with the external field. The difference between the current density $\textbf{J}_{1}$ and $\textbf{J}_{2}$ at $f_{q-BIC}$ leads to the radiation towards the free space, exhibiting as a dark mode -- a discrete eigenmode that manifests as a sharp resonances with clear background spectrum. The transition from in-phase to out-of-phase current oscillations at $f_{q-BIC}$ emanates from the singular nature of BIC. By breaking the non-radiating nature of the ordinary oscillations, an abrupt energy accumulated in the current oscillations can be emitted into the free space, with high Q factor and plain surroundings. In addition, by managing the asymmetry degree of the unit cells, the resonance frequency and the quality factor of the BIC-based resonance mode can be tuned, which will be presented in the following discussions.

\section{Method of Solution and Discussion of Results}\label{method}

\subsection{Formation of quasi bound states in the continuum}

The opposite-phase current oscillations illustrated in the last section is an intuitive indicator of BIC-based resonance mode, which is associated with the singularity of BICs. Besides judging the quasi-BIC resonance mode by the direct way of observing current distributions, in this part, via an improved eigenmode analysis, we will further demonstrate the formation of quasi BICs in our proposed structures without symmetry breaking ($\Delta L = \Delta h = 0$~nm).

Due to the large imaginary dielectric permittivity of gold ($\varepsilon_{gold}^{\prime\prime}$), here we manage to attenuate the influence of intrinsic metallic losses, and decrease the imaginary part of permittivity by orders of magnitude. Thus, with the reduced imaginary permittivity of the system $\varepsilon^{\prime\prime}$, we could have a clear look at the Q factors and eigenfrequencies in the band structure, which is the typical approach to verify the existence of BICs.

\begin{figure}[!htb]
\centering
\includegraphics[width=0.95\columnwidth]{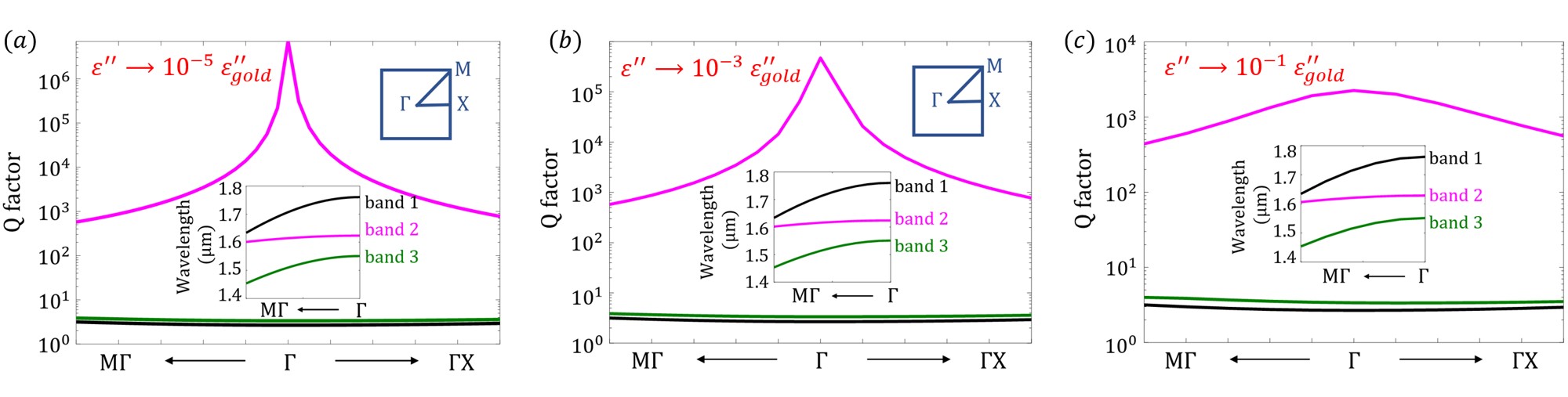}
\caption{Q factors and band structures (insets) of symmetric nanostructures ($\Delta L = \Delta h = 0$~nm) with decreased metallic losses, with the imaginary part of permittivity $\varepsilon^{\prime\prime}$ is reduced to (a) $10^{-5} \varepsilon_{gold}^{\prime\prime}$ (b) $10^{-3} \varepsilon_{gold}^{\prime\prime}$ and (c) $10^{-1} \varepsilon_{gold}^{\prime\prime}$, respectively. The quasi BICs we studied appears at the $\Gamma$ point of the first Brillouin zone, with the Q factors exceeding (a) $10^{6}$ (b) $10^{5}$ and (c) $10^{3}$, respectively.}
\label{fig_Q}
\end{figure}

With dispersive metallic permittivity, the tranditional eigenmode analysis in COMSOL Multiphysics cannot give the accurate eigen solution for our proposed structure, because the permittivity of the system is not clear under an unknown eigenfrequency. Here we use an improved eigenmode analysis which involves with the stationary
solver with an initial solution (details are demonstrated in Appendix B).

Fig. 4 presents the calculated Q factors in $M \Gamma$ and $\Gamma X$ directions, with different $\varepsilon^{\prime\prime}$ [the first Brillouin zone of the square lattice is plotted on the upper right of Fig.~4(a)].
When the imaginary part of permittivity $\varepsilon^{\prime\prime}$ is reduced to (a) $10^{-5} \varepsilon_{gold}^{\prime\prime}$ (b) $10^{-3} \varepsilon_{gold}^{\prime\prime}$ and (c) $10^{-1} \varepsilon_{gold}^{\prime\prime}$, the corresponding Q factors would be greater than (a) $10^{6}$ (b) $10^{5}$ and (c) $10^{3}$ for band 2 (marked with carmine curve), protected by $\Gamma$ point. The inset figures at the bottom of (a)-(c) show all the possible eigenmodes around 1.6$\mu$m in the $M\Gamma$ direction. For band 1 and band 3, corresponding Q factors are less than 10, whereas band 2 exhibits the representative performance of quasi BICs -- calculated Q factor rises up rapidly to an extremely large value at the symmetry point $\Gamma$. Note that, when the imaginary part of permittivity is set to be 0, the quasi-BIC becomes genuine BIC, and thus at the BIC point, the radiative Q factor will diverge to infinity due to the absence of loss. In other words, the total Q value is limited by the losses in the system. Since practical applications are limited to the quasi-BIC regime, in the following discussions, we will use the actual permittivities of metals to study the linear and nonlinear optical responses of the proposed metasurfaces.

\subsection{Quantum and classical explanation of quasi BIC in plasmon-MoS2 metasurfaces}

Having proved that the eigenmode around 1.6$\mu$m emanates from quasi BIC, in this part, we will further explain the quasi BIC mode from the perspective of quantum and classical optics, respectively.

Firstly, beginning with the famous Fano-Anderson Hamilton which addresses the existence of BIC, we make a quantum analogue for our proposed structure and illustrate the formation of quasi BICs. For simplicity, here we only consider a single unit cell rather than the whole metasurface. For complex meta-structures in reality, the conclusions can be understood similarly from this example. As described in Fig. 5, the light-trapping meta-atom made of gold strips is analogous to a discrete level with state $| a \rangle$, and the surroundings made up of free space, monolayer MoS2, and substrate below are analogized by a common continuum with state $| k \rangle$. All the decay channels can be analogized as the coupling interaction between the discrete state and the continuum.

\begin{figure}[!htb]
\centering
\includegraphics[width=0.6\columnwidth]{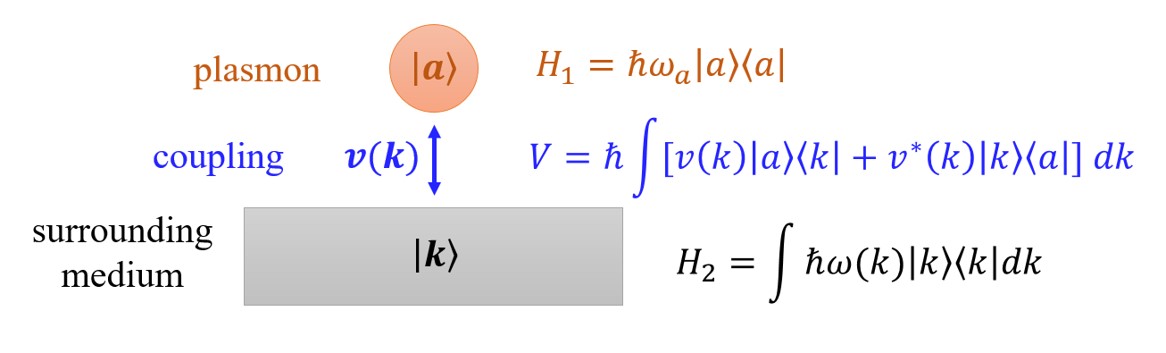}
\caption{Proposed Fano-Anderson-like model\cite{eaa18book, s09jmo, ay05pre, assy05pre, s07epjb} to describe the system involving with a single localized state coupled to a surrounding continuum. The upper meta-atom made of gold strips is analogized as a defect mode with single discrete level (discrete state $| a \rangle$), the surrounding medium made up of MoS2, air and substrate is analogized as a common continuum (continuous state $| k \rangle$), and the coupling interaction is the quantum analogue for energy decay [coupling coefficient $\mathbf{\upsilon}(k)$].  The Hamiltonian of the non-interacting discrete and continuous sates, and the interacting part is described by $H_{1}$, $H_{2}$ and $V$, respectively.}
\label{schematic}
\end{figure}

The model in Fig. 5 is described by the Hamiltonian

\begin{equation}
\label{eq:H_sum}
H=\hbar\omega_{a}| a \rangle \langle a | + \int \hbar\omega (k) | k \rangle \langle k | dk + \hbar\int[ v(k) | a \rangle \langle k | + v^{\ast}(k) | k \rangle \langle a | ] dk
\end{equation}
where $H_{1}=\hbar\omega_{a}| a \rangle \langle a |$ and $H_{2}=\int \hbar\omega (k) | k \rangle \langle k | dk$ are the Hamiltonian of the non-interacting discrete (localized) state $| a \rangle$ and continuous states $| k \rangle$, respectively, and $V=\hbar\int[ v(k) | a \rangle \langle k | + v^{\ast}(k) | k \rangle \langle a | ] dk$ is the interacting part. Representing the wave function of the system as $| \phi \rangle = a(t)| a \rangle + \int b(k,t)| k \rangle dk$, the coupled-mode equations of the system go as follows, with the amplitude expansion coefficient $a(t)$ for state $| a \rangle$, and $b(k,t)$ for $| k \rangle$, respectively,

\begin{align}
\label{eq:couple_time}
i \dot{a}(t) &= \omega_{a} a(t)+ \int v(k)b(k,t) dk \\
i \dot{b}(k,t) &= \omega (k)b(k,t)+v^{\ast}(k)a(t)
\end{align}
where the overdot stands for the derivative in time. At the eigenstates of the system, $ H | \phi_{E} \rangle = \hbar \Omega | \phi_{E} \rangle $, $a(t) = \tilde{a} e^{-i\Omega t}$, $b(k,t) = \tilde{b}(k) e^{-i\Omega t}$, by introducing the density of states $\rho (\omega) = \partial k / \partial \omega$, and substituting $a(t)$, $b(k,t)$ into Eq.(2) and (3), we obtain:

\begin{align}
\label{eq:couple_frequency}
\Omega \tilde{a} & = \omega_{a} \tilde{a} + \int \rho (\omega) v (\omega)\tilde{b}(\omega) d\omega \\
\Omega \tilde{b}(\omega) & = \omega \tilde{b}(\omega) +v^{\ast}(\omega)\tilde{a}
\end{align}
Using the equations above, the link between the amplitude of discrete mode and continuous mode can be written as

\begin{equation}
\label{eq:link}
\tilde{a}=\frac{\int \rho(\omega)v(\omega) \tilde{b}(\omega) d \omega}{\Omega - \omega_{a}}
\end{equation}

The above result corresponds to the following physics: (i) when the eigenfrequency of the system $\Omega$ coincides with the localized energy of the discrete mode $\omega_{a}$, the amplitude of the discrete state $ \tilde{a}$ becomes infinitely large; (ii) for realistic cases where the amplitude $ \tilde{a}$ is limited, the coupling coefficient between the discrete state and continuous state $v(\omega)$ would be close to zero, implying that the localized state interacts negligibly with the continuum and exhibits the destructive interference effect\cite{ay05pre, assy05pre, s07epjb}.

Besides the quantum explanation as illustrated above, the quasi BICs of our proposed metasurfaces can also be explained via the classical optics analysis. In order to investigate classical optical correspondence of the system, the dispersion maps of reflectance with respect to the asymmetric parameters ($\Delta L$ and $\Delta h$) are plotted as shown in Fig.~6(a,c). We observe the Fano resonance dip of reflection around the optical communication wavelength, which originates from the physics of BICs as a result of distortion of the in-plane (a) and out-of-plane (c) symmetry-protected BIC. The reflection dip occurs at the eigenfrequencies of the system, and it vanishes when the geometry becomes symmetric in both cases. In the context of in-plane asymmetric metasurfaces [Fig. 6(a)], the quasi-BIC resonance has a small frequency displacement to the short-wavelength band with increasing asymmetric parameter $\Delta L$. As $\Delta L$ increases, the bandwidth broadens, leading to lower Q factors (which we will clarify in detail later in this subsection). In the case of out-of-plane asymmetric metasurfaces [Fig.~6(c)], when $\Delta h$ increases, the quasi-BIC trapped mode resonance is shifted towards the long-wavelength band, which can be utilized for tuning the optical resonance frequencies.

\begin{figure}[!htb]
\centering
\includegraphics[width=0.95\columnwidth]{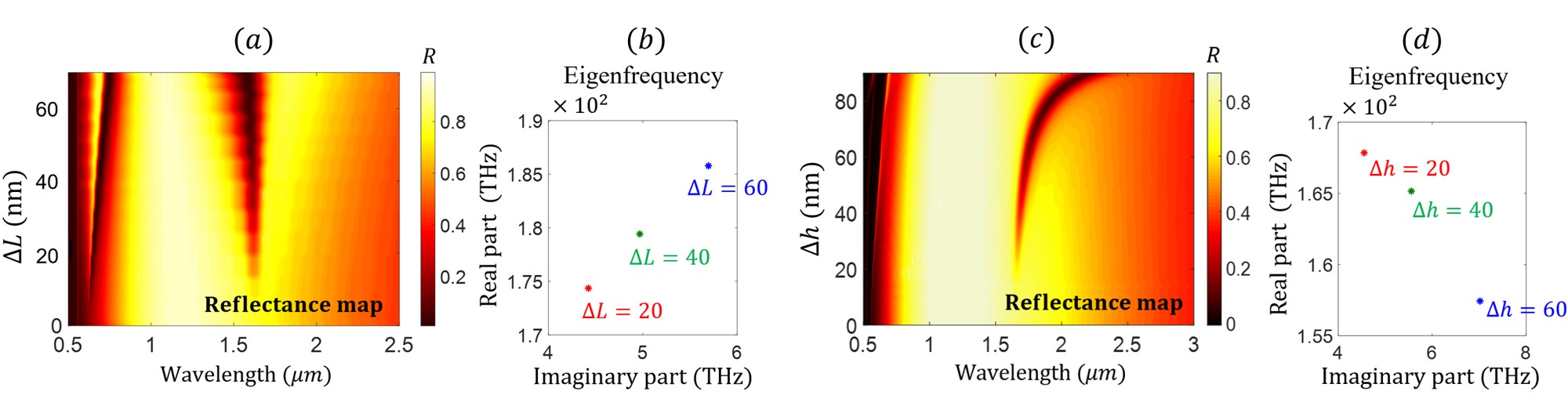}
\caption{(a,c) Dispersion map of reflection for (a) in-plane and (c) out-of-plane symmetry-broken metausurfaces with respect to the asymmetric parameters $\Delta L$ and $\Delta h$, respectively. Parameters are (a) $L_{y1}=(120 - \Delta L)$~nm, $L_{y2}=(120 + \Delta L)$~nm, $w = 50$~nm, and $h_{gold} = 100$~nm (c) $L_{y1}=L_{y2}=120$~nm, $w = 50$~nm, $h_{gold1}=(100 + \Delta h)$~nm, and $h_{gold2}= (100 - \Delta h) $~nm. (b,d) Complex eigenfrequencies at $\Gamma$ point in band structure for (b) in-plane and (d) out-of-plane symmetry-broken meta atoms. Each point represents the eigenmode of the hybrid meta-atom, calculated for different asymmetric parameters (b) $\Delta L$ and (d) $\Delta h$, plotted with real part on the vertical axis and imaginary part on the horizontal.  }
\label{schematic}
\end{figure}

The reflection maps in Fig.~6 also show the differences between plasmon-polariton resonances and quasi-BIC resonances. In symmetric systems when $\Delta L =0$ ($\Delta h =0$), the former one (ordinary plasmon-polariton resonance) always exists in visible spectrum range, while the latter one (quasi-BIC-based trapped mode resonance) disappears within the infrared wavelength region. By introducing the asymmetric factor ($\Delta L$, $\Delta h$) either in plane or out of plane, a light-trapping mode resonance can be formed in the infrared spectrum wavelength with a clean background.

In addition, different from a true BIC achieved in ideal lossless and infinite structures (closed system) with real eigenmode frequency and vanishing resonance width, in lossy metals (open systems), BICs collapse into quasi-BICs with complex eigenfrequencies. The eigenmode analysis (for details see Appendix B) we applied for calculating eigenfrequencies stems from the eigen equation (Maxwell's equation in discrete Fourier space): $\nabla\times (\nabla\times \textbf{E})-k_{0}^{2}\varepsilon_{r}\textbf{E}=0$, where the field $E$ is in the time-harmonic representation, $\textbf{E}(\textbf{r},t) = Re(\textbf{E}(\textbf{r})e^{-i\omega t}) $, which includes a complex parameter $\omega$ in the phase - the real part represents the eigenfrequency and the imaginary part is responsible for losses. It can be seen in Fig. 6(b, d) that the eigenfrequencies for in-plane and out-of-plane asymmetric metasurfaces have various complex eigenfrequencies with different asymmetric parameters $\Delta L$ $(\Delta h)=20, 40, 60$~nm. The real parts and imaginary parts of the eigenmode frequencies are responsible for the resonance positions (trapped mode frequency $f_{q-BIC}$) and linewidths (damping) of corresponding reflectance maps, respectively. The Q factor of the system can be derived from the eigenmode frequencies, approximate to the ratio between real part and imaginary part of eigenfrequencies\cite{ksmak18prl}. There is a variation of Q factors when the unit-cell asymmetry changes - specifically, the Q factor of in-plane (out-of-plane) symmetry-broken meta-atoms decreases when the asymmetric parameter $\Delta L$ ($\Delta h$) gets larger.

In the context of in-plane symmetry-broken meta-atoms, the real part of eigenfrequency gets larger when $\Delta L$ increases [Fig. 6(b)], corresponding to the results in Fig. 6(a) - the optical resonance shifts to the short-wavelength band with increasing asymmetric parameter $\Delta L$. In addition, the imaginary part of eigenfrequency also gets larger with increasing $\Delta L$, which corresponds to broader bandwidth with larger $\Delta L$ illustrated in Fig. 6(a). The Q factors are about 54, 33, 24 for $\Delta L = 20, 40, 60$~nm, respectively, calculated from Fig. 6(b). Under the circumstance when the symmetry of meta-atom is broken out of plane, the real part of eigenfrequency decreases when $\Delta h$ increases [Fig. 6(d)], which corresponds to the result in Fig. 6(c) - the optical resonance has a displacement to the long-wavelength band with larger $\Delta h$. Moreover, the imaginary part increases when $\Delta h$ increases, which also agrees with Fig. 6(c) - the bandwidth broadens with increasing asymmetric degree. In this case, the Q factors are approximately 36, 29, 22 for $\Delta h = 20, 40, 60$~nm, respectively, calculated from Fig. 6(d). Normally due to the large damping rates ($\gamma\sim 10^{13}-10^{14}$~Hz) of metals \cite{eaa18book}, the propagating plasmons in infrared band decay very quickly. Nevertheless, the BIC-inspired Fano resonances shown in Fig. 6(a,c) could clearly increase the lifetime of plasmon oscillations (magnitude of picosecond), with the imaginary part of eigenfrequency being around $10^{12}$~Hz magnitude. Therefore, with enhanced lifetime of plasmons, the BIC-based Fano resonances would give rise to an accumulation of field intensity at metal surfaces (hot spots), which could be adopted for further enhancing the nonlinear optical processes.

\begin{figure}[!htb]
\centering
\includegraphics[width=0.95\columnwidth]{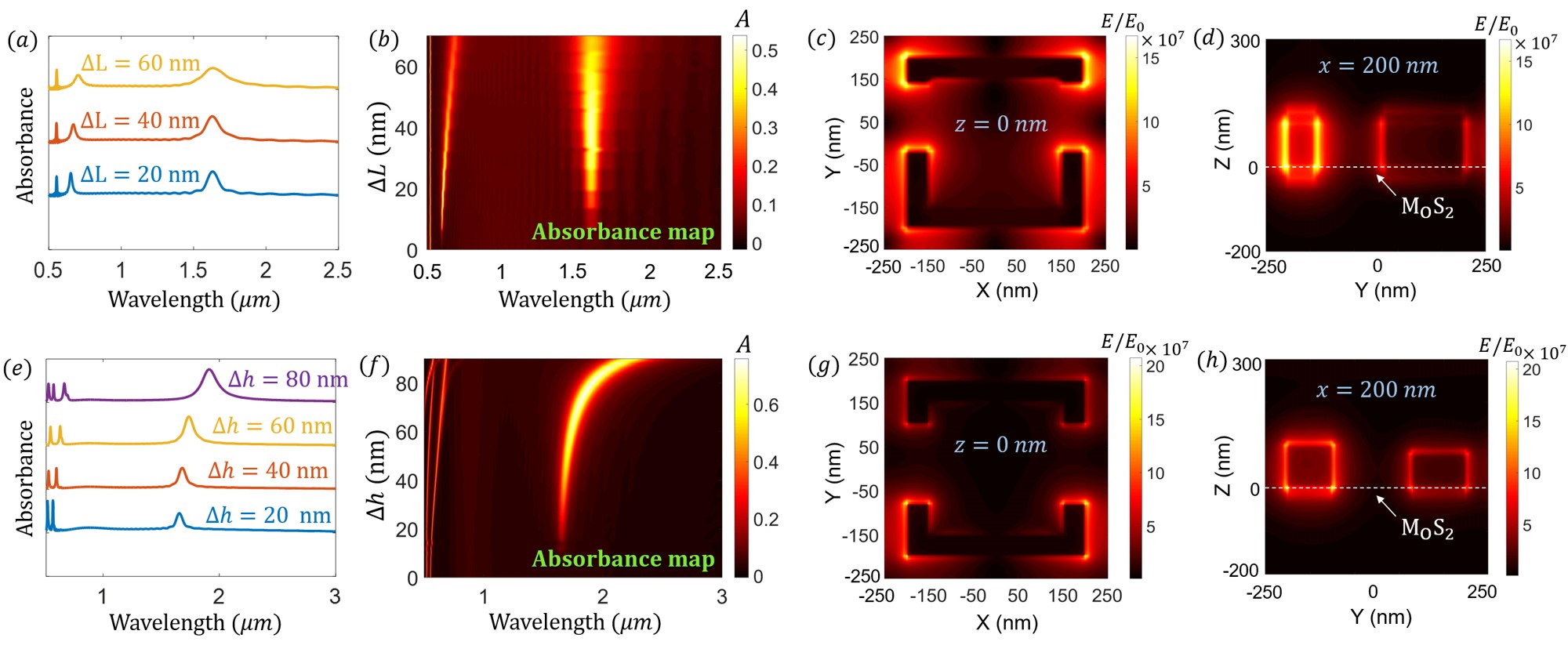}
\caption{(a,e) Evolution of the absorbance spectra with respect to (a) in-plane and (e) out-of-plane asymmetric parameter $\Delta L$ and $\Delta h$, respectively. Parameters are the same as the structure described in Fig. 5. The bandwidth broadens with increasing asymmetric parameter. (b,f) Dispersion map of absorbance vs (b) in-plane and (f) out-of-plane antisymmetric parameter $\Delta L$ and $\Delta h$. (c, d, g, h) Cross section of electric field mode profile distribution at (c,g) $z=0$~nm and (d,h) $x=200$~nm for the two generic metasurfaces at $f_{q-BIC}$. [(c,d) - in plane; (g,h) - out of plane symmetry breaking] Enhancement of E field can be up to $10^{8}$ at hot spots as displayed in the colorbar.}
\label{schematic}
\end{figure}

\subsection{Enhanced nonlinear optical channels in communication band}

The enhancement of optical nonlinearity mainly comes from two sources: The first source is the near-field localization of the incident radiation, which could enhance the local field intensity by as large as 5 orders of magnitude \cite{ma12np}. Localized modes inspired by plasmons could enhance the nonlinear processes via boosting the linear optical response. It can be seen in Fig. 7(a,e) that, by introducing an asymmetric element either in plane or out of plane, an appreciable absorption occurs in the infrared band, which also signifies large enhancement of light at infrared wavelength. Moreover, the bandwidth of absorbance broadens with increasing asymmetric parameters, which corresponds to the analysis of Q factors in Sec.III-B -- larger asymmetric degree makes Q factor decreases whereas damping increases, and thus widening the bandwidth of the optical response.

In addition to the field localization, another factor which improves the nonlinear optical process in infrared band is the Fano resonance brought by in-plane or out-of-plane symmetry breaking in geometry. In accordance with the reflectance map presented above, the dispersion maps of absorbance of in-plane and out-of-plane symmetry-broken metasurfaces are given in Fig. 7(b) and Fig. 7(f), respectively. In contrast to the reflectance spectrum which exhibits dips at resonances, the absorbance spectrum displays peaks, with a frequency shift under the same geometrical parameter $\Delta L$ ($\Delta h$). The frequency shift between absorbance peaks and reflectance dips would introduce another Fano response in the transmission spectrum, which enhances the localized field furthermore.

On account of the above factors, the E field at the resonance mode $f_{q-BIC}$ can be enhanced by up to $10^{8}$ orders of magnitude as compared to the incident field $E_{0}$, which is also indicated in Fig. 7(c,d,g,h). Inside the metal strips, the field is negligible, due to the field shielding effect of metal. However, at the interface between metal and dielectric (air/MoS2), there is ultra intensive localized near field driven by field confinement of plasmons and BIC-based Fano responses. With such a strong localization of the electromagnetic field at the hot spots, the nonlinear response of frequency converting materials (monolayer MoS2 in our case) positioned near hot spots can be greatly enhanced.

With intrinsically broken crystal inversion symmetry and large second-order nonlinear responses, monolayers of transition-metal dichalcogenides such as MoS2 have shown great promise for future nonlinear light sources \cite{ltaka13prb, hvadm17lsa, fmh19pn}. With the field enhancement effect stated above, the second harmonic generation efficiency of MoS2 can be boosted furthermore, via placing on hot spots. To measure the enhancement of the optical nonlinearity of frequency-conversion systems, an important physical quantity - nonlinear susceptibility tensor, is taken into account \cite{ama18mt, qun19prb}. We firstly describe the electric field of the generated second harmonic light $\textbf{E}(2\omega)$ in terms of the second-order susceptibility $\chi^{(2)}$ and input light $\textbf{E}(\omega)$:

\begin{equation}
\label{eq:chi}
\textbf{E}(2\omega)=\varsigma\chi^{(2)}:\textbf{E}(\omega)\textbf{E}(\omega)
\end{equation}
where $\omega$ is the input signal frequency, $2\omega$ is the second harmonic frequency and $\varsigma$ denotes a proportionality coefficient that contains local field factors determined by the local dielectric environment. For single layer MoS2 with $D_{3h}$ point group symmetry, the second-harmonic susceptibility tensor has only one nonvanishing element: $\chi^{(2)}_{MoS2}\equiv \chi^{(2)}_{xxx} = -\chi^{(2)}_{xyy} = -\chi^{(2)}_{yyx} = -\chi^{(2)}_{yxy}$ , thus calculating one element of susceptibility tensor $\chi^{(2)}_{xxx}$ is adequate for analyzing second harmonic generation (SHG) efficiency \cite{r08book,y03book}. Utilizing the homogenization method, the effective nonlinear response of the monolayer MoS2 can be derived from the spatial overlap integral between fields within the structure by probing light at fundamental modes with specific polarization combinations averaged over the monolayer MoS2 \cite{qun19prb, qun18oe, qun20access, jedn18josa, mjjn16prb}:

\begin{equation}
\label{eq:chi_eff}
\bm{\chi}^{(2)}_{eff,xxx} = \frac{\frac{1}{V}\int_{V}\chi^{(2)}_{xxx} (\textbf{r})E_{x} (\textbf{r}) E_{x}(\textbf{r}) d\textbf{r}^{3}}{\frac{1}{V}\int_{V}E_{x}(\textbf{r})d\textbf{r}^{3} \cdot \frac{1}{V}\int_{V}E_{x}(\textbf{r})d\textbf{r}^{3} }
\end{equation}
where $E_{x}$ is the x component of the near field excited in MoS2, and $V$ is the volume of monolayer MoS2. Thus the enhancement of second-order susceptibility in MoS2 can be represented as

\begin{equation}
\label{eq:eta}
\eta = \int_{V} VE_{x} (\textbf{r}) E_{x}(\textbf{r}) d\textbf{r}^{3} / \left[\int_{V}E_{x}(\textbf{r})d\textbf{r}^{3} \right]^{2}
\end{equation}

Calculations of $\chi^{(2)}$ enhancement in MoS2 are summarized in Fig.~8(a) and (b), for the in-plane and out-of-plane asymmetric metasurfaces, respectively. One important result inferred from the presented data is that, a remarkable enhancement of second-order nonlinearity occurs within the plasmon resonance region. In particular, the maximum SHG efficiency enhancement can be up to 125 and 81, corresponding to in-plane, and out-of-plane asymmetric structures, respectively. With increasing asymmetric parameters $\Delta L$ and $\Delta h$, the peak values of $\chi^{(2)}$ enhancement decrease, for these two generic metasurfaces, respectively. Since $\chi^{(2)}$ around $f_{PP}$ can also be enhanced by orders of magnitude through the field localization effect, the nonlinear susceptibilities of the metasurfaces can be improved at different wavelength channels, which provides a promising route for high-efficiency wavelength-multiplexed nonlinear optical signal processing that could be integrated on a nanoscale structure.

Moreover, it can also be observed that differs from the absorbance spectra in Fig.~7(a,e), the spectra of SHG nonlinearity enhancement show a Fano-like behavior -- exhibiting both enhancement and suppression phenomena for the SHG process, which was also observed in the experiments \cite{jgmpg12oe}. The suppresion phenomenon can be simply interpreted by the fact that the transparency fuelled by BIC-based Fano resonance does not permit excitation on conversion state, essentially due to destructive interference effect in the linear response. To give a rigid mathematical explanation, a quantum model is set up as follows, for the SHG process in our system.


\begin{figure}[!htb]
\centering
\includegraphics[width=0.95\columnwidth]{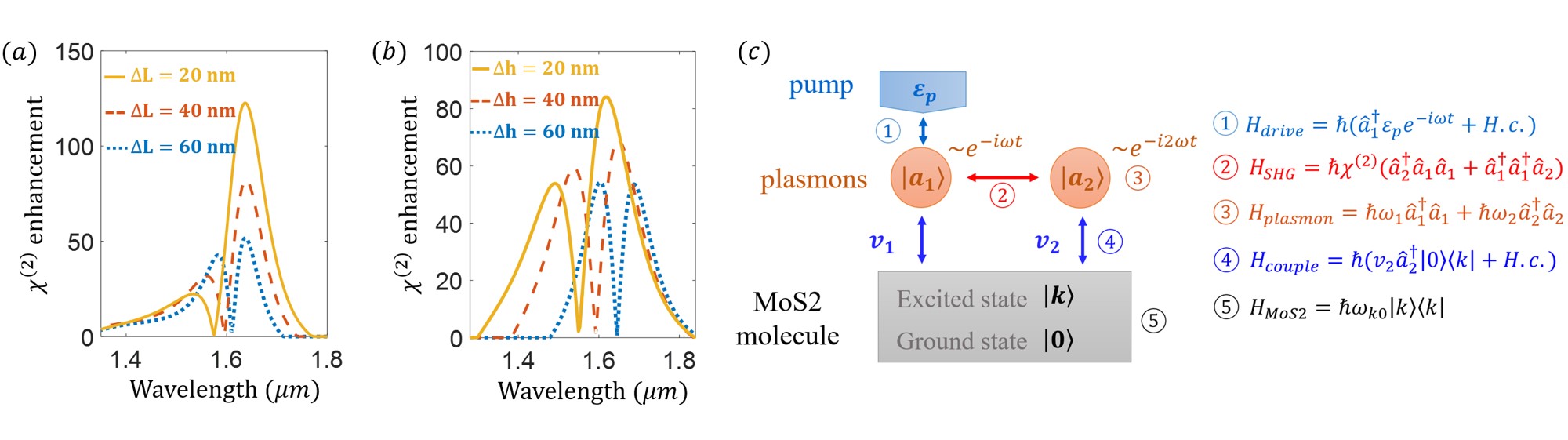}
\caption{(a, b) Wavelength dependence of the enhancement of second-order susceptibility in MoS2 determined for the (a) in-plane, and (b) out-of-plane asymmetric plasmon structures on top of MoS2, with various asymmetric parameters $\Delta L$ and $\Delta h$,  respectively. (c) Schematic model of the proposed plasmon-MoS2 system, with the SHG process. The two-way arrows mark all the coupling interactions in the system, which includes the interaction of the plane wave source with plasmon, plasmon with molecule, and fundamental plasmon mode with second-order plasmon mode. $|a_{1}\rangle$ and $|a_{2}\rangle$ denote the fundamental and second-order modes of plasmons, respectively, $|k\rangle$ and $|0\rangle$ denote the excited state and ground state of MoS2 molecules, respectively. $\varepsilon_{p}$ in $H_{drive}$ is a coefficient proportional to the amplitude of driving plane wave.}
\label{schematic}
\end{figure}

Described in Fig. 8(c), we present an analogic quantum model for the SHG process. An incident wave with pump frequency $\omega$ couples to the fundamental plasmon mode $|a_{1}\rangle$ with oscillations $e^{-i\omega t}$, and creates a second-order mode $|a_{2}\rangle$ with frequency $2\omega$ oscillating with $e^{-i2\omega t}$. MoS2 molecules are positioned at the hot spots, leading to strong coupling interaction with plasmons. Here since we only consider the SHG-associated effect, the coupling between molecule and fundamental plasmon mode $|a_{1}\rangle$ is ignored (namely $v_{1}=0$) in the following discussions.

Beginning with our analysis from the Hamiltonian which signifies all the dynamic processes within the system, the Hamiltonian of the system can be represented as the sum of the five terms: $H_{drive}$ (the interaction of the incident pump with the fundamental plasmon mode $|a_{1}\rangle$), $H_{SHG}$ (the SHG process involving with modes $|a_{1}\rangle$ and $|a_{2}\rangle$), $H_{plasmon}$ (the energy of plasmons), $H_{couple}$ (the interaction of the second harmonic generated plasmons with MoS2 molecule), and $H_{MoS2}$ (the energy of MoS2 molecules). Applying the Heisenberg equation and replacing $|a \rangle$ with corresponding complex amplitude $a$ as performed in Sec.~III-B, the equations of motion (EOM) in time domain can be attained as follows:

\begin{align}
\label{eq:SHG_time}
\dot{a_{1}} & = (-i\omega_{1} - \gamma_{1})a_{1} - i2\chi^{(2)}a_{1}^{\ast}a_{2} + \varepsilon_{p}e^{-i\omega t} \\
\dot{a_{2}} & = (-i\omega_{2}- \gamma_{2})a_{2} - i\chi^{(2)}a_{1}^{2} - iv_{2}\rho_{0k} \\
\dot{\rho_{0k}} & = (-i\omega_{k0}-\gamma_{k0})\rho_{0k} + iv_{2}a_{2}(\rho_{kk} - \rho_{00}) \\
\dot{\rho_{kk}} & = -\gamma_{kk}\rho_{kk} + iv_{2}a_{2}^{\ast}\rho_{0k} - ia_{2}\rho_{0k}^{\ast}
\end{align}
where the decay rates of plasmons $\gamma_{1,2}$ and molecules $\gamma_{kk}$ are introduced, corresponding to the imaginary element of complex eigenmodes. The typical damping rates for molecules $\gamma_{kk}$ are approximate to $10^{12}$~Hz, and around $10^{12}$~Hz for plasmons in our case \cite{mm17aop, m13nano, cmp02ajp, ppl06prl, dgam14jop}. For steady-state eigenmodes, switching the time-domain EOM to the frequency domain, via taking the forms $a_{1}(t)=\tilde{a}_{1}e^{-i\omega t}$, $a_{2}(t)=\tilde{a}_{2}e^{-i2\omega t}$, $\rho_{0k}(t)=\tilde{\rho}_{0k}e^{-i2\omega t}$, $\rho_{kk}(t)=\tilde{\rho}_{kk}$, the frequency-domain EOM are

\begin{align}
\label{eq:SHG_frequency}
[i(\omega_{2}-2\omega) & +\gamma_{2}]\tilde{a}_{2}+i\chi^{(2)}\tilde{a}_{1}^{2}=-iv_{2}\tilde{\rho}_{0k} \\
[i(\omega_{k0}-2\omega)& +\gamma_{k0}]\tilde{\rho}_{0k}=iv_{2}\tilde{a}_{2}(\tilde{\rho}_{kk}-\tilde{\rho}_{00})
\end{align}
eliminating the irrelevant variable $\tilde{\rho}_{0k}$ in the above equations, we can find the relationship between SHG generated modes and fundamental modes, that is

\begin{equation}
\label{relate}
\tilde{a}_{2}=\frac{i\chi^{(2)}}{\frac{v_{2}^{2}(\tilde{\rho}_{kk}-\tilde{\rho}_{00})} {i(\omega_{k0}-2\omega)+\gamma_{k0}}-[i(\omega_{2}-2\omega)+\gamma_{2}]}\tilde{a}_{1}^{2}
\end{equation}

Turning our attention to the denominator of the $\tilde{a}_{2}$ vs. $\tilde{a}_{1}$ relationship equation, the first term $v_{2}^{2}(\tilde{\rho}_{kk} - \tilde{\rho}_{00})/\gamma_{k0}$ could reach an ultra large value on resonance $\omega_{k0}=2\omega$. The underlying reason is that when excited on resonance, the molecule exhibits extremely narrow linewidth as compared to the unexcited states, making $\gamma_{k0}$ extremely small. Thus, the ultra largeness of $v_{2}^{2}(\tilde{\rho}_{kk}-\tilde{\rho}_{00})/\gamma_{k0}$ dominates the denominator of this relationship equation, resulting in the suppression phenomenon in the SHG process presented in Fig.~8(c). Importantly, the suppression effect on optical nonlinearity can be used for prohibiting undesired conversion process at certain wavelength region, thus bringing in a narrower linewidth.

\subsection{Suggested Fabrication Process}

The nanostructure discussed in the paper can be fabricated by the process shown in Fig. 9. The process can be basically separated into two parts: preparation of MoS2 monolayers [Fig. 9(a)-(b)] and fabrication of asymmetric metal structures [Fig. 9(c)-(j)]. The MoS2 monolayer can be synthesized by CVD method. The source of S and Mo comes from sulfur and molybdenum oxide powder, respectively, which could be put into fused quartz tubes of the CVD furnace. The growth of MoS2 could be carried out under atmospheric pressure and the protection of argon carrier gas. Fabrication of metal structures with different heights (out-of-plane asymmetric structure) can be implemented via the e-beam lithography (EBL) for twice. For instance, in Fig. 9(c)-(j), Au films with different heights are deposited on the MoS2 monolayer by two independent EBL, with two lift-off processes afterwards. While for in-plane asymmetric structure, only one EBL and one lift-off are needed.

\begin{figure}[!htb]
\centering
\includegraphics[width=0.95\columnwidth]{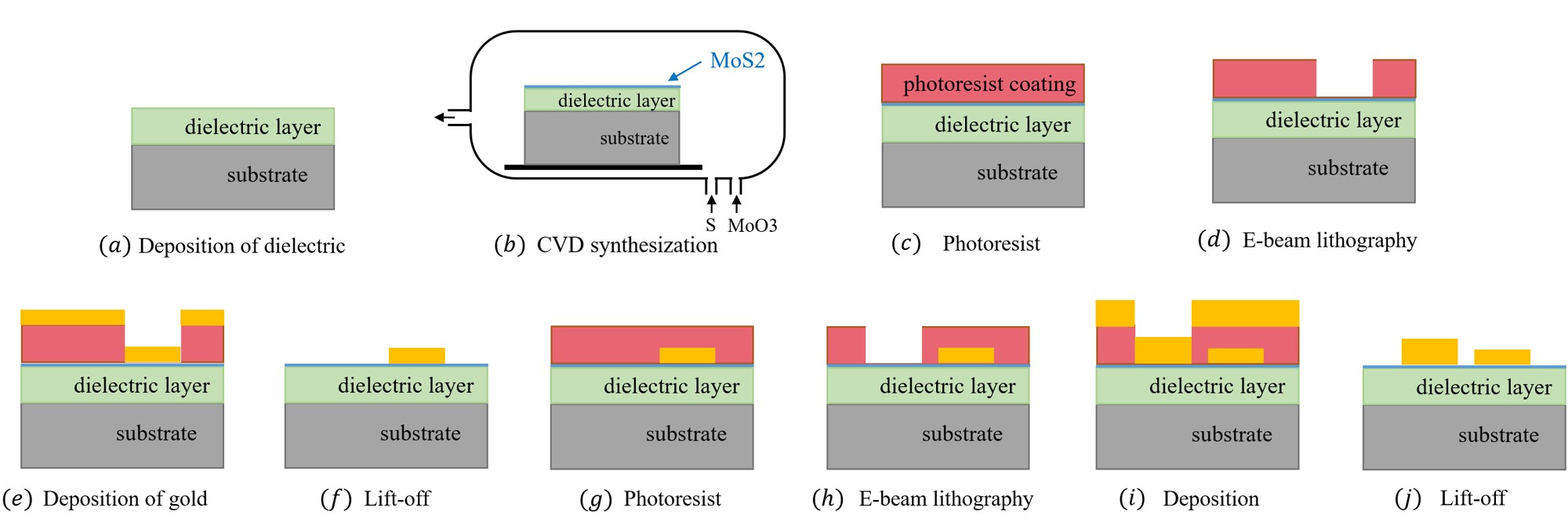}
\caption{Suggested fabrication process, including (a-b) the preparation of MoS2 and (c-j) the fabrication of asymmetric metal structures. Structures with different heights can be accomplished by the photoresist coating, EBL, deposition as well as the lift-off processes for twice.}
\label{schematic}
\end{figure}

\section{Conclusion}\label{Concl}

In summary, we have proposed a multiplexing-oriented plasmon-assisted MoS2 metasurface with superior SHG efficiency. Via in-plane and out-of plane symmetry breaking in the meta atom, a BIC-based radiating modes can be excited in optical communication band with high Q factors. The out-of-plane asymmetric structure presents frequency tunability as compared to the in-plane asymmetry case, by shifting asymmetric parameters. In addition, we have also demonstrated the difference between the BIC-based resonance and the ordinary plasmon-polariton resonance, from the perspective of both classical and quantum mechanisms. Moreover, utilizing the well-known homogenization approach, we further computed the SHG enhancement of MoS2 metasurfaces driven by the quasi-BIC mode, and demonstrated that the enhancement of second-order nonlinear susceptibility can be up to 2 orders of magnitude, which is promising for bringing in extra SHG channels for wavelength-multiplexed nonlinear applications.

We also analyzed the Fano-like phenomenon for optical nonlinearity. A suppression of nonlinear response occurs for in-plane and out-of-plane metasurfaces proposed in this study, driven by the destructive interference effect. Also, a quantum model was proposed to explain the reason of suppression. In the end of this work, we gave a suggested fabrication procedure for the asymmetric structures we proposed, which may facilitate relevant experimental verifications of our proposal. Finally, extension of our ideas to topological photonics and ultrafast laser optics may be promising for realizing highly integrated optical networks \cite{zhihao20prl,zhihao20prb,mnn2019, nmn2020}.

\section*{Acknowledgments}\label{Ack}

Dr. Qun Ren sincerely thank Prof. Nicolae-Coriolan Panoiu for his seasoned supervising. We acknowledge the financial support by National Program on Key Basic Research Project of China (Grant No. 2019YFB2203602), National Natural Science Foundation of China (Grant No. 61975149) and Independent Innovation Fund of Tianjin University (Grant No. 2020XYF-0111).

%
%
%
%
%
%
%
\clearpage
\setcounter{figure}{0}
\renewcommand{\thefigure}{A\arabic{figure}}

\appendix



\section{Current distribution of gold strips}

For the numerical simulations of the current distribution of gold strips, we used the finite-element-method solver under the Electromagnetic Waves, Frequency Domain (ewfd) interface in COMSOL Multiphysics. All calculations are realized for a 500~nm $\times$ 500~nm unit cell of gold-MoS2 hybrid structure on SiO$_{2}$ substrate (height of 500~nm, refractive index $n=1.5$). The geometry sizes of gold strips were described in the main text in Sec.~II. Material properties for gold was imported from the tabulated data in the reference work \cite{at08apb} mentioned in main text. Corresponding complex permittivity of gold is shown as below:

\begin{figure}[!htb]
\centering
\includegraphics[width=0.45\columnwidth]{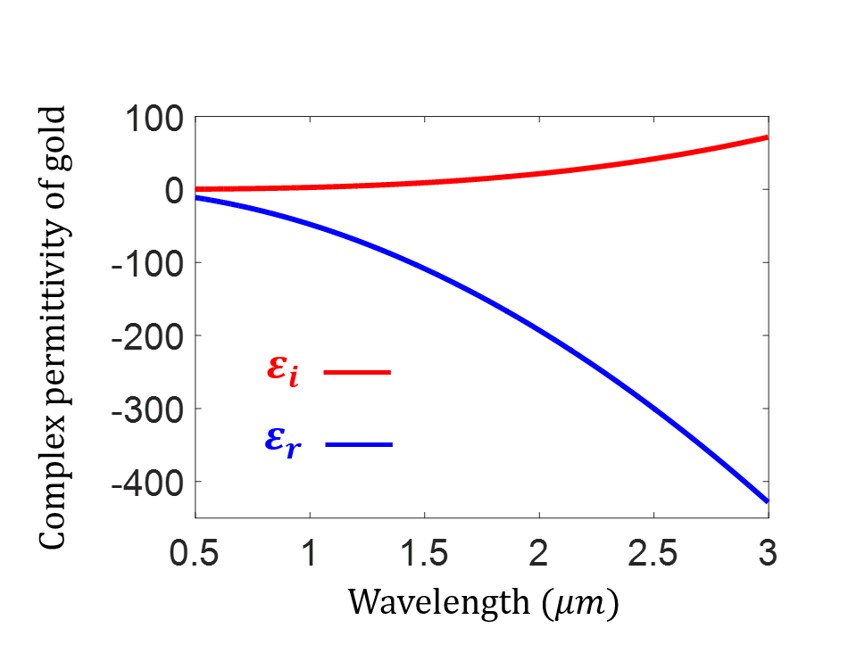}
\caption{Real part (blue curve) and imaginary part (red curve) of the dielectric function of gold from the operating wavelength 500~nm to 3000~nm. The imaginary part is responsible for the loss, and the real part for the degree of polarization of the medium to the external electric field.}
\label{A1}
\end{figure}

Setting with Floquet periodic boundary conditions and one tenth of operating wavelength for the maximum cell size in free space, the optical response of the system was swept from wavelength 0.5~$\mu$m to 3~$\mu$m. Specifically, the relationship between the electric current density $\textbf{J}$, the magnetic field intensity $\textbf{H}$, and the electric displacement field $\textbf{D}$ is given as follows:

\begin{equation}
\label{eq:current}
\nabla\times \textbf{H} = \frac{\partial \textbf{D}}{\partial t} + \textbf{J} \Rightarrow \textbf{J}=\nabla\times \textbf{H}-\frac{\partial \textbf{D}}{\partial t} \Rightarrow \textbf{J}=
\left|                 
  \begin{array}{ccc}   
    \textbf{i} & \textbf{j} & \textbf{k}\\  
    \frac{\partial}{\partial x} & \frac{\partial}{\partial y} & \frac{\partial}{\partial z}\\  
    H_{x} & H_{y} & H_{z}\\
  \end{array}
\right|
-(-i\omega \textbf{D})
\end{equation}
Setting the three components of electric current density vector as

\begin{align}
\label{eq:Jx_Jy_Jz}
J_{x} & = \frac{\partial H_{z}}{\partial y} - \frac{\partial H_{y}}{\partial z} +i\omega D_{x} \\
J_{y} & = \frac{\partial H_{x}}{\partial z} - \frac{\partial H_{z}}{\partial x} +i\omega D_{y}  \\
J_{z} & = \frac{\partial H_{y}}{\partial x} - \frac{\partial H_{x}}{\partial y} +i\omega D_{z}
\end{align}
Thus the amplitude of the current density can be expressed as $J=\sqrt{J_{x}^{2} + J_{y}^{2}+J_{z}^{2} }$, and the arrows which indicate the direction of current can be determined via the three components of the current density vector: $J_{x}$, $J_{y}$, and $J_{z}$.

Sweeping the current distribution from 0.5~$\mu$m to 3~$\mu$m, we find that the opposite-phase current oscillations occur for the quasi-BIC mode [Fig.~3(a,b)], whereas the in-phase current oscillations occur for the ordinary plasmon-polariton resonance modes [Fig.~2(a,b)]. Current distributions on either side of the quasi-BIC mode $f_{q-BIC}$ are shown in Fig. A2 below.

\begin{figure}[!htb]
\centering
\includegraphics[width=0.8\columnwidth]{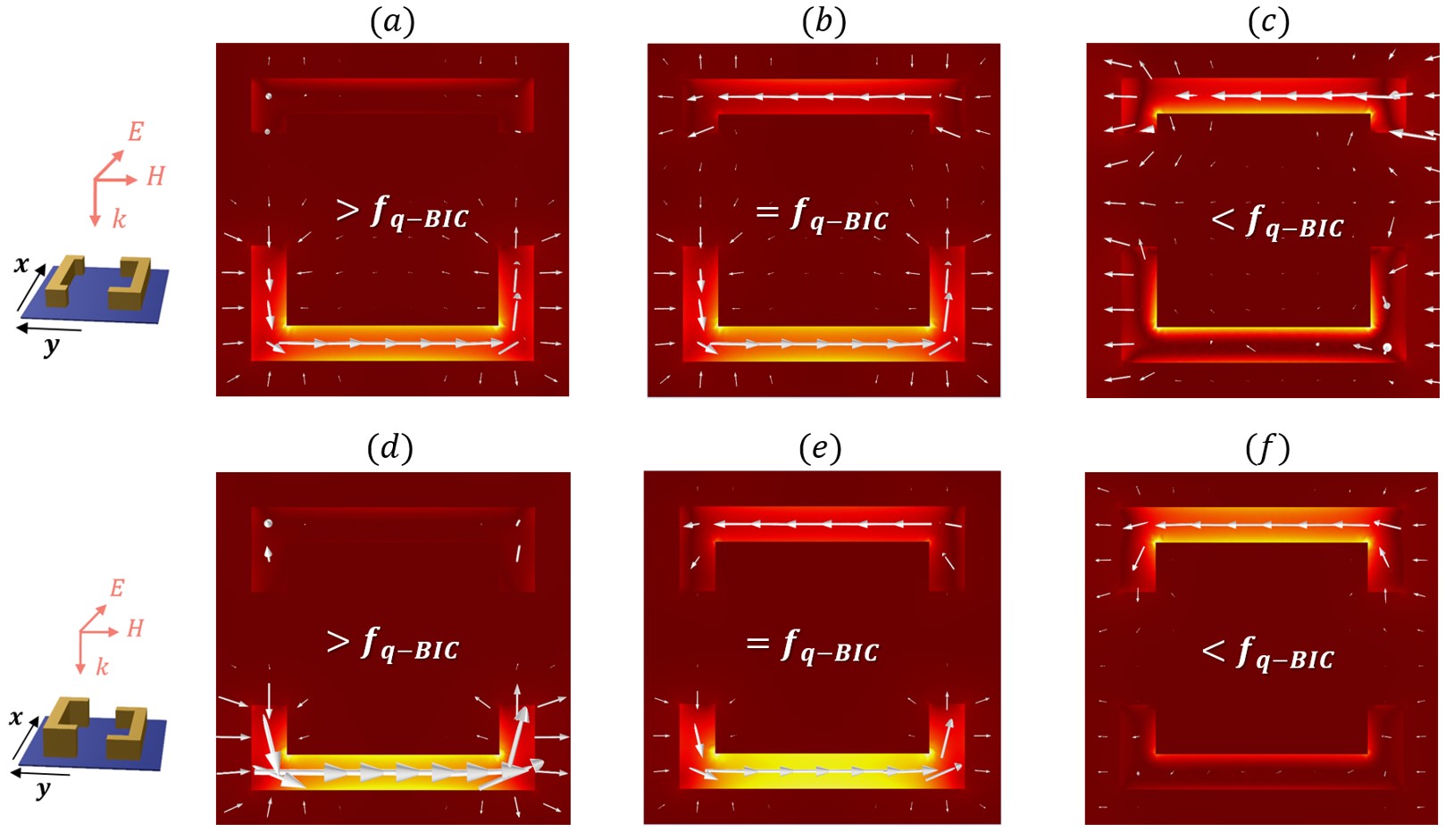}
\caption{Evolution of current distribution around the quasi-BIC mode $f_{q-BIC}$ for (a-c) in-plane and (d-f) out-of-plane symmetry-broken metasurfaces.}
\label{A2}
\end{figure}

For in-plane and out-of-plane asymmetric metasurfaces, there exist transition states for the current distribution [Fig. A2(a,c) and (d,f)] between the quasi-BIC mode (anti-phase current oscillations) and ordinary radiation modes (in-phase current oscillations) of metal structures -- the current oscillations disappear on one side of the gold strips around $f_{q-BIC}$. This results further confirms the singularity of the quasi-BIC mode of the metasurfaces we proposed in the main text.

\section{Band structure and Q factor of BIC-based metamaterials}

For the numerical calculations of the Q factor and band structure (especially eigenfrequency) of our proposed nanostructure, we employed eigenfrequency solver under the ewfd interface in COMSOL. The model was built in three dimensions, with a unit cell made of gold and MoS2. The Floquet periodic conditions are set both in $x$ and $y$ directions, and the perfectly matched layers (PML) are constructed in the $z$ direction.

\begin{figure}[!htb]
\centering
\includegraphics[width=0.9\columnwidth]{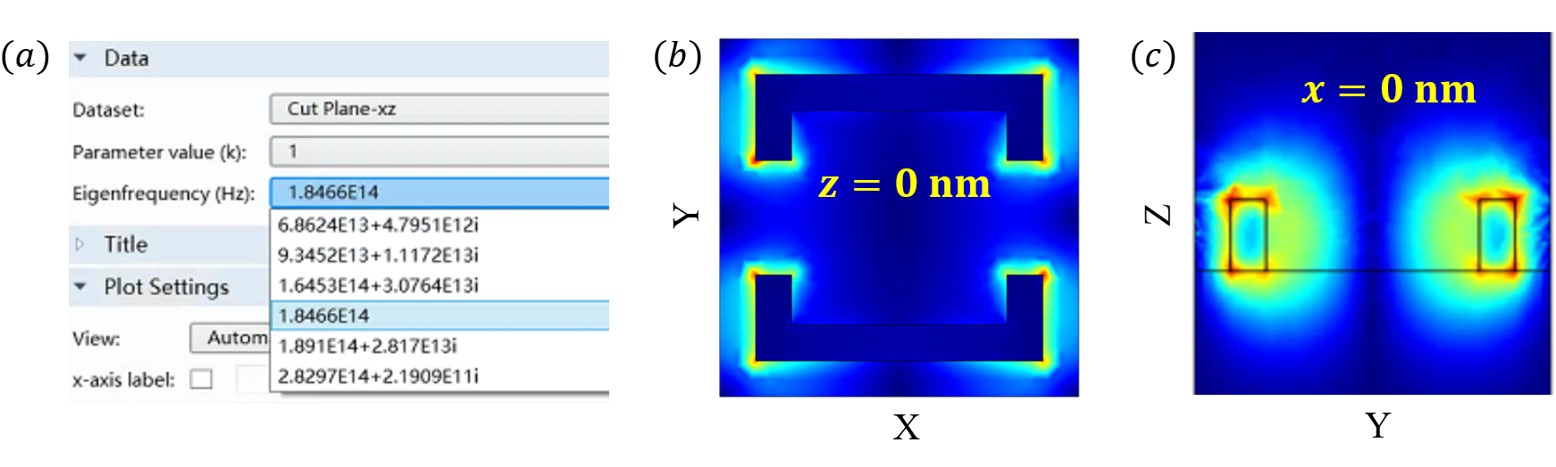}
\caption{(a) Calculated eigenfrequencies (including both true and false modes) of our proposed system via COMSOL eigenmode analysis; (b,c) E field distribution at the cut plane when (b) $z=0$~nm and (c) $x=0$~nm under the BIC-based eigenfrequency ($f=1.8466\times 10^{14}$~Hz in this case).}
\label{A3}
\end{figure}

Due to the dispersive property of the system (the permittivity of gold depends on the frequency), the permittivity is not clear under an unknown eigenfrequency, thus the typical procedure of applying eigenfrequency solver cannot give the right solution. Here the simulation is divided into three steps: (i) For study 1 (ewfd), get the initial solution of eigenfrequency setting with a certain reference value of eigenfrequency; (ii) Set up the constraint condition via ODE function with an integral operator normalizing the E field of the system to ; (iii) For study 2 (ewfd2), utilizing the stationary solver to get the self-consistent solution under the global constraint set in step (ii) (similar approach had also been illustrated in one case from COMSOL application library, but in a simple 2D system -- `bandgap-photonic-crystal').

As mentioned in the main text, in order to verify the formation of quasi-BIC mode in the infrared region, we managed to decrease the imaginary part of permittivity of gold, and thus mitigated the effects of intrinsic metallic losses on the Q factors of the proposed system. Take an instance when imaginary part of permittivity equals to $10^{-5}\times \varepsilon^{''}_{gold}$, even though there is still little loss in the system, the eigenfrequency at $\Gamma$ point is real, whereas other `false' eigenmodes calculated by COMSOL all have complex eigenfrequencies [Fig.~A3(a)]. This is also another way for judging the true/false eigenmode in COMSOL (the other way is simply to compare the Q factors of these modes).

In addition, to double check our proposed approach of tuning the imaginary part of metallic permittivity is correct, we also plotted the E field profiles [at Fig.~A3(b) $z=0$~nm and (c) $x=0$~nm] under the eigenfrequency $f=1.8466\times 10^{14}$~Hz and $k=\Gamma$. We find that the field distributions on $x$-$y$ and $y$-$z$ plane are the same as the real cases (simulations with actual permittivity of gold). Tuning the imaginary part of gold permittivity would not affect the typical characteristics of metals - localize electromagnetic field on the surface of metals and enhance the localized field by several orders of magnitude.

\section{Accuracy of homogenization method}

The theoretical method we utilized to extract the effective optical coefficients (especially the effective nonlinear susceptibility) is the improved homogenization technique demonstrated in our previous work \cite{qun19prb, qun18oe, jedn18josa}. Our proposed metasurfaces can be considered as homogenized metasurfaces with effective physical quantities, such as effective permittivities and effective susceptibilities.

\begin{figure}[!htb]
\centering
\includegraphics[width=0.75\columnwidth]{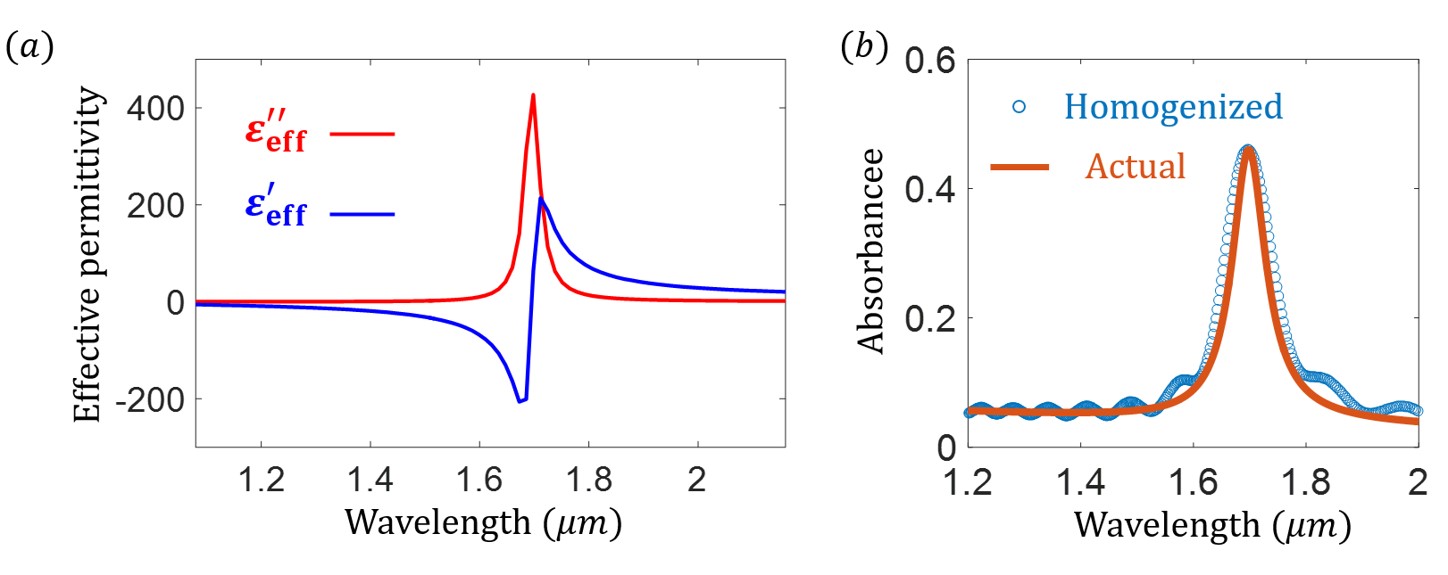}
\caption{(a) Effective relative permittivity of the proposed metasurfaces via homogenization approach, with $L_{y1}=L_{y2}=120$~nm, $w=50$~nm and $\Delta h=40$~nm. The real and imaginary part are plotted with blue and red curve, respectively. (b) Absorbance comparison calculated for the actual metasurface (depicted with solid curve) and its homogenized counterpart (marked with circles) with the effective relative permittivity $\varepsilon_{eff}$ plotted in (a).}
\label{A4}
\end{figure}

To validate the accuracy of the homogenization we applied in the main text, here we compare the optical response (absorbance) of the homogenized metasurfaces with the actual metasurfaces with proposed structures. Based on the constitutive relation of materials, the electric displacement $\textbf{D}$ and electric field $\textbf{E}$ is expressed as

\begin{equation}
\label{eq:D_E}
D_{i}=\sum_{j}\varepsilon_{ij}E_{j}
\end{equation}
where the subscripts $i, j=x, y, z$. Thus the averaged fields of the metasurface can be introduced as

\begin{align}
\label{eq:D_E_eff}
\textbf{D}_{eff}(\omega) & = \frac{1}{V}\int_{V}\mathbf{D}(\mathbf{r},\omega)d\mathbf{r} \\
\textbf{E}_{eff}(\omega) & = \frac{1}{V}\int_{V}\mathbf{E}(\mathbf{r},\omega)d\mathbf{r}
\end{align}
where $V$ is the volume of the unit cell of the metasurface. Thus, utilizing the above equations, the effective electric permittivity of the metasurface can be defined as

\begin{equation}
\label{eq:eps_eff}
\mathbf{\varepsilon}_{eff}(\omega)=\frac{\int_{V}\mathbf{D}(\mathbf{r}, \omega)d\mathbf{r}}{\int_{V}\mathbf{E}(\mathbf{r},\omega)d\mathbf{r}} = \frac{\int_{V}\varepsilon (\textbf{r})\mathbf{E}(\mathbf{r}, \omega)d\mathbf{r}}{\int_{V}\mathbf{E}(\mathbf{r},\omega)d\mathbf{r}}
\end{equation}

Calculating the effective permittivity given by Eq.~\ref{eq:eps_eff}, we find that the effective permittivity of the homogenized metasurface presents an evident resonance response at the quasi-BIC mode [Fig.~A4(a)], which also verifies the optical enhancing influence brought by metallic plasmons. To assess the validity of the homogenization approach we employed in this work, a reliable way is to compare the optical response (absorbance for instance) between the homogenized structure and the original one. Summarized in Fig.~A4(b), the absorbance of the homogenized structure agrees well with the actual metasurface, thus proving the accuracy this homogenization approach in our computations. Note that, for this accuracy test of homogenization, we just used limited number of frequency monitors, leading to weak oscillations in the smooth region of absorbance. When the frequency monitors and meshes in unit cell set in simulations are more enough, the absorbance in these two cases will agree better. Anyway, for the purpose of getting the peak value of effective parameters, the current settings in the simulations are sufficient for us to get the correct results.


\begin{thebibliography}{0}%
\makeatletter
\providecommand \@ifxundefined [1]{%
 \@ifx{#1\undefined}
}%
\providecommand \@ifnum [1]{%
 \ifnum #1\expandafter \@firstoftwo
 \else \expandafter \@secondoftwo
 \fi
}%
\providecommand \@ifx [1]{%
 \ifx #1\expandafter \@firstoftwo
 \else \expandafter \@secondoftwo
 \fi
}%
\providecommand \natexlab [1]{#1}%
\providecommand \enquote  [1]{``#1''}%
\providecommand \bibnamefont  [1]{#1}%
\providecommand \bibfnamefont [1]{#1}%
\providecommand \citenamefont [1]{#1}%
\providecommand \href@noop [0]{\@secondoftwo}%
\providecommand \href [0]{\begingroup \@sanitize@url \@href}%
\providecommand \@href[1]{\@@startlink{#1}\@@href}%
\providecommand \@@href[1]{\endgroup#1\@@endlink}%
\providecommand \@sanitize@url [0]{\catcode `\\12\catcode `\$12\catcode
  `\&12\catcode `\#12\catcode `\^12\catcode `\_12\catcode `\%12\relax}%
\providecommand \@@startlink[1]{}%
\providecommand \@@endlink[0]{}%
\providecommand \url  [0]{\begingroup\@sanitize@url \@url }%
\providecommand \@url [1]{\endgroup\@href {#1}{\urlprefix }}%
\providecommand \urlprefix  [0]{URL }%
\providecommand \Eprint [0]{\href }%
\providecommand \doibase [0]{http://dx.doi.org/}%
\providecommand \selectlanguage [0]{\@gobble}%
\providecommand \bibinfo  [0]{\@secondoftwo}%
\providecommand \bibfield  [0]{\@secondoftwo}%
\providecommand \translation [1]{[#1]}%
\providecommand \BibitemOpen [0]{}%
\providecommand \bibitemStop [0]{}%
\providecommand \bibitemNoStop [0]{.\EOS\space}%
\providecommand \EOS [0]{\spacefactor3000\relax}%
\providecommand \BibitemShut  [1]{\csname bibitem#1\endcsname}%
\let\auto@bib@innerbib\@empty
\end{thebibliography}%


\begin{thebibliography}{99}


\bibitem{mcr04ol} 
M. Zimmermann, C. Gohle, R. Holzwarth, T. Udem, and T. W. Hansch,``Optical clockwork with an offset-free difference-frequency comb: accuracy of sum- and difference-frequency generation,'' Opt. Lett. \textbf{29}, 310–312, (2004).


\bibitem{xkm17light} 
M. Xin, K. Safak, M. Y. Peng, A. Kalaydzhyan, W. Wang, O. D. Mucke and F. X. Kartner, ``Attosecond precision multi-km laser-microwave network,'' Light Sci. Appl. \textbf{6}, e16187 (2017).

\bibitem{mcpl17n} 
M. Kues, C. Reimer, P. Roztocki, L. Romero Cortes, S. Sciara, B. Wetzel, Y. Zhang, A. Cino, S. T. Chu, B. E. Little, D. J. Moss, L. Caspani, J. Azana, and R. Morandotti, ``On-chip generation of high-dimensional entangled quantum states and their coherent control,'' Nature \textbf{546}, 622-626 (2017).


\bibitem{mzw19lpr}
M. Ma, Z. Li, W. Liu, C. Tang, Z. Li, H. Cheng, J. Li, S. Chen, and J. Tian, ``Optical Information Multiplexing with Nonlinear Coding Metasurfaces,'' Laser Photonics Rev. \textbf{13}, 1900045 (2019).


\bibitem{mnw19prl} 
M. Fang, N. H. Shen, Wei E. I. Sha, Z. Huang, T. Koschny, and C. M. Soukoulis, ``Nonlinearity in the dark: broadband terahertz generation with extremely high efficiency,'' Phys. Rev. Lett. \textbf{122}, 027401 (2019).

\bibitem{tx17pr}
T. Wang and X. Zhang, ``Improved third-order nonlinear effect in graphene based on bound states in the continuum,'' Photonics Res. \textbf{5}, 629-639 (2017).


\bibitem{lr19aom}
L. Cong and R. Singh, ``Symmetry‐Protected Dual Bound States in the Continuum in Metamaterials,'' Adv. Opt. Mater. \textbf{7}, 1900383 (2019).


\bibitem{jrqn18nano}
J. W. You, S. R. Bongu, Q. Bao, and N. C. Panoiu, ``Nonlinear optical properties and applications of 2D materials: theoretical and experimental aspects,'' Nanophotonics \textbf{8}, 63-97 (2018).

\bibitem{gwx19oe}
G. Y. Chen, W. X. Zhang, and X. D. Zhang, ``Strong terahertz magneto-optical phenomena based on quasi-bound states in the continuum and Fano resonances,'' Opt. Express \textbf{27}, 16449-16460 (2019).

\bibitem{hh19pr}
H. J. Lee and H. S. Park, ``Generation and measurement of arbitrary four-dimensional spatial entanglement between photons in multicore fibers,'' Photonics Res. \textbf{7}, 19-27 (2019).

\bibitem{bqs19lpr}
B. Guo, Q. L. Xiao, S. H. Wang, and H. Zhang, ``2D layered materials: synthesis, nonlinear optical properties, and device applications,'' Laser Photonics Rev. \textbf{13}, 1800327 (2019).


\bibitem{tbc19pra}
T. Guo, B. Jin, and C. Argyropoulos, ``Hybrid graphene-plasmonic gratings to achieve enhanced nonlinear effects at terahertz frequencies,'' Phys. Rev. Appl.	\textbf{11}, 024050 (2019).

\bibitem{ysqy20oe}
Y. Liu, S. Zhu, Q. Zhou, Y. Cao, Y. Fu, L. Gao, H. Chen and Y. Xu, ``Enhanced third-harmonic generation induced by nonlinear field resonances in plasmonic-graphene metasurfaces,'' Opt. Express \textbf{28}, 13234-13242 (2020).

\bibitem{kyk19acs}
K. Koshelev, Y. Tang, K. Li, D. Y. Choi, G. Li, and Y. Kivshar, ``Nonlinear metasurfaces governed by bound states in the continuum,'' ACS Photonics \textbf{6}, 1639-1644 (2019).

\bibitem{lei20lpr}
Y. Meng, Q. Zhang, D. Lei, Y. Li, S. Li, Z. Liu, W. Xie, and C. W. Leung, ``Plasmon‐Induced Optical Magnetism in an Ultrathin Metal Nanosphere‐Based Dimer‐on‐Film Nanocavity,'' Laser Photonics Rev. 2000068 (2020).




%


\bibitem{jzn20sa}
J. W. You, Z. Lan, and N. C. Panoiu, ``Four-wave mixing of topological edge plasmons in graphene metasurfaces,'' Sci. Adv. \textbf{6}, eaaz3910 (2020).

\bibitem{nwd18jo} 
N. C. Panoiu, Wei E. I.Sha, D. Y. Lei, and G. C. Li, ``Nonlinear optics in plasmonic nanostructures,'' J. Opt. \textbf{20}, 083001 (2018).




\bibitem{aa18jo}
A. Krasnok and A. Alu, ``Embedded scattering eigenstates using resonant metasurfaces,'' J. Opt. \textbf{20}, 064002 (2018).

\bibitem{fd75pra}
F. H. Stillinger and D. R. Herrick, ``Bound states in the continuum,'' Phys. Rev. A \textbf{11}, 446 (1975).




\bibitem{xuyi2020pra}
Y. Lin, T. Feng, S. Lan, J. Liu, and Y. Xu, ``On-Chip Diffraction-Free Beam Guiding beyond the Light Cone,'' Phys. Rev. Appl. \textbf{13}, 064032 (2020).



\bibitem{ksmak18prl}
K. Koshelev, S. Lepeshov, M. Liu, A. Bogdanov, and Y. Kivshar, ``Asymmetric metasurfaces with high-Q resonances governed by bound states in the continuum,'' Phys. Rev. Lett. \textbf{121}, 193903 (2018).


\bibitem{bfdk5ome} 
B. Mukherjee, F. Tseng, D. Gunlycke, K. K. Amara, G. Eda, and E. Simsek, ``Complex electrical permittivity of the monolayer molybdenum disulfide (MoS2) in near UV and visible,'' Optical Materials Express \textbf{5}, 447-455 (2015).

\bibitem{ywj18oe} 
Y. Jiang, W. Chen, and J. Wang, ``Broadband MoS2-based absorber investigated by a generalized interference theory,'' Opt. Express \textbf{26}, 24403-24412 (2018).


\bibitem{at08apb} 
A. Vial and T. Laroche, ``Comparison of gold and silver dispersion laws suitable for FDTD simulations,'' Appl. Phys. B \textbf{93}, 139–43 (2008).

\bibitem{eaa18book}
E. Kamenetskii, A. Sadreev, and A. Miroshnichenko, ``Fano Resonances in Optics and Microwaves,'' New York, Springer (2018).

\bibitem{nsnv08np}
N. I. Zheludev, S. L. Prosvirnin, N. Papasimakis, and V. A. Fedotov, ``Lasing spaser,'' Nat. photonics \textbf{2}, 351-354 (2008).


\bibitem{ay05pre}
A. E. Miroshnichenko and Y. S. Kivshar, ``Engineering Fano resonances in discrete arrays,'' Phys. Rev. E \textbf{72}, 056611 (2005).

\bibitem{assy05pre}
A. E. Miroshnichenko, S. F. Mingaleev, S. Flach, and Y. S. Kivshar, ``Nonlinear Fano resonance and bistable wave transmission,'' Phys. Rev. E \textbf{71}, 036626 (2005).

\bibitem{s07epjb}
S. Longhi, ``Bound states in the continuum in a single-level Fano-Anderson model,'' Eur. Phys. J. B \textbf{57}, 45-51 (2007).
%
%
%


\bibitem{ma12np}
M. Kauranen and A. V. Zayats, ``Nonlinear plasmonics,'' Nat. Photon. \textbf{6}, 737-748 (2012).


%
%
%
%
%
%
%



\bibitem{ltaka13prb}
L. M. Malard, T. V. Alencar, A. P. M. Barboza, K. F. Mak, and A. M. De Paula, ``Observation of intense second harmonic generation from MoS2 atomic crystals,'' Phys. Rev. B \textbf{87}, 201401 (2013).

\bibitem{hvadm17lsa}
H. Chen, V. Corboliou, A. S. Solntsev, D. Y. Choi, M. A. Vincenti, D. De Ceglia, C. de Angelis, Y. Lu, and D. N. Neshev, ``Enhanced second-harmonic generation from two-dimensional MoSe2 on a silicon waveguide,'' Light Sci. Appl. \textbf{6}, e17060-e17060 (2017).

\bibitem{fmh19pn}
F. A. Zarif, M. K. Nezhad, and H. R. M. Rezaieun, ``Enhancement of efficiency of second-harmonic generation from MoS2 monolayers in 1D Fibonacci photonic crystals,'' Photonics Nanostruct. Fundam. Appl. \textbf{36}, 100726 (2019).

\bibitem{ama18mt}
A. Krasnok, M. Tymchenko, and A. Alu, ``Nonlinear metasurfaces: a paradigm shift in nonlinear optics,'' Mater. Today \textbf{21}, 8-21 (2018).


\bibitem{r08book}
R. Boyd, Nonlinear optics, Academic Press, London, (2008).

\bibitem{y03book}
Y. R. Shen, The Principles of Nonlinear Optics, Wiley Interscience, New York, (2003).


\bibitem{qun19prb}
Q. Ren, J. W. You, and N. C. Panoiu, ``Large enhancement of the effective second-order nonlinearity in graphene metasurfaces,'' Phys. Rev. B \textbf{99}, 205404 (2019).

\bibitem{qun18oe}
Q. Ren, J. W. You and N. C. Panoiu, ``Giant enhancement of the effective Raman susceptibility in metasurfaces made of silicon photonic crystal nanocavities,'' Opt. Express \textbf{26}, 30383-30392 (2018).

\bibitem{qun20access}
Q. Ren, J. W. You, and N. C. Panoiu, ``Comparison between the linear and nonlinear homogenization of graphene and silicon metasurfaces,'' IEEE Access, 10.1109/ACCESS.2020.3026313 (2020).

\bibitem{jedn18josa}
J. W. You, E. Threlfall, D. F. Gallagher, and N. C. Panoiu,  "Computational analysis of dispersive and nonlinear 2D materials by using  a GS-FDTD method," J. Opt. Soc. Am. B \textbf{35}, 2754-2763 (2018).



%
%

\bibitem{mjjn16prb}
M. Tymchenko, J. S. Gomez-Diaz, J. Lee, N. Nookala, M. A. Belkin, and A. Alu, ``Advanced control of nonlinear beams with Pancharatnam-Berry metasurfaces,'' Phys. Rev. B \textbf{94}, 214303 (2016).

\bibitem{mm17aop}
M. Premaratne and M. I. Stockman, ``Theory and technology of spasers,'' Adv. Opt. Photon. \textbf{9}, 79–128 (2017).

\bibitem{m13nano}
M. E. Tasgin, ``Metal nanoparticle plasmons operating within a quantum lifetime,'' Nanoscale \textbf{5}, 8616–8624 (2013).

\bibitem{cmp02ajp}
C. L. Garrido Alzar, M. A. G. Martinez, and P. Nussenzveig, ``Classical analog of electromagnetically induced transparency,'' Am. J. Phys. \textbf{70}, 37–41 (2002).

\bibitem{ppl06prl}
P. Anger, P. Bharadwaj, and L. Novotny, ``Enhancement and quenching of single-molecule fluorescence,'' Phys. Rev. Lett. \textbf{96}, 113002 (2006).


\bibitem{dgam14jop}
D. Turkpence, G. B. Akguc, A. Bek, and M. E. Tasgin, ``Engineering nonlinear response of nanomaterials using Fano resonances,'' J. Opt. \textbf{16}, 105009 (2014).



\bibitem{jgmpg12oe}
J. Berthelot, G. Bachelier, M. Song, P. Rai, G. C. Des Francs, A. Dereux, and A. Bouhelier, ``Silencing and enhancement of second-harmonic generation in optical gap antennas,'' Opt. Express \textbf{20}, 10498-10508 (2012).


\bibitem{zhihao20prl}
Z. Lan, J. W. You, Q. Ren, Wei E. I. Sha, and N. C. Panoiu, ``Second-harmonic generation via double topological valley-Hall kink modes in all-dielectric photonic crystals,'' arXiv preprint arXiv:2007.04875 (2020).

\bibitem{zhihao20prb}
Z. Lan, J. W. You, and N. C. Panoiu, ``Nonlinear one-way edge-mode interactions for frequency mixing in topological photonic crystals,'' Phys. Rev. B \textbf{101}, 155422 (2020).

\bibitem{mnn2019}
M. Xin, N. Li, N. Singh, A. Ruocco, Z. Su, E. Salih Magden, J. Notaros, D. Vermeulen, E. P. Ippen, M. R. Watts, and F. X. Kaertner, ``Optical frequency synthesizer with an integrated erbium tunable laser,'' Light Sci. Appl. \textbf{8}, 122 (2019).


\bibitem{nmn2020}
N. Singh, M. Xin, N. Li, D. Vermeulen, A. Ruocco, E. S. Magden, K. Shtyrkova, E. P. Ippen, F. X. Kaertner, and M. R. Watts, ``Silicon Photonics Optical Frequency Synthesizer,'' Laser Photonics Rev. 1900449 (2020).


\end{thebibliography}
\end{document}